\input epsf
\openup 1 pt
\magnification\magstephalf
\overfullrule 0pt
\voffset -1.5truecm
\vsize 24.5 truecm
\def\gsim{\raise.3ex\hbox{$\;>$\kern-.75em\lower1ex\hbox{$\sim$}$\;$}}
\def\P{I\!\!P}
\font\rfont=cmr10 at 10 true pt
\def\ref#1{$^{\hbox{\rfont {[#1]}}}$}

\font\helv=phvr8r scaled 975
\font\bold=phvb8r scaled 975
\font\oblique=phvro8r scaled 975

\font\tenbfit=cmbxti10
\font\sevenbfit=cmbxti10 at 7pt
\font\fivebfit=cmbxti10 at 5pt
\newfam\bfitfam 
\textfont\bfitfam=\tenbfit  \scriptfont\bfitfam=\sevenbfit
\scriptscriptfont\bfitfam=\fivebfit

\font\tenbfit=cmbxti10
\font\sevenbfit=cmbxti10 at 7pt
\font\fivebfit=cmbxti10 at 5pt
\newfam\bfitfam 
\textfont\bfitfam=\tenbfit  \scriptfont\bfitfam=\sevenbfit
\scriptscriptfont\bfitfam=\fivebfit

\font\tenbit=cmmib10
\newfam\bitfam
\textfont\bitfam=\tenbit%

\font\tenmbf=cmbx10
\font\sevenmbf=cmbx7
\font\fivembf=cmbx5
\newfam\mbffam
\textfont\mbffam=\tenmbf \scriptfont\mbffam=\sevenmbf
\scriptscriptfont\mbffam=\fivembf

\font\tenbsy=cmbsy10
\newfam\bsyfam 
\textfont\bsyfam=\tenbsy%


\def\pmb#1{\setbox0=\hbox{#1}
 \kern.05em\copy0\kern-\wd0 \kern-.025em\raise.0433em\box0 }

\def\slash{/\kern-.5em}


 %


\def\boxit#1{\vbox{\hrule\hbox{\vrule\kern1pt\vbox
{\kern1pt#1\kern1pt}\kern1pt\vrule}\hrule}}

\def\h{\hfill\break}
\parskip=6pt
\parindent=0pt
\hsize=17truecm\hoffset=-5truemm
\def\footnoterule{\kern-3pt
\hrule width 17truecm \kern 2.6pt}


\catcode`\@=11 

\def\nolabels{\def\wrlabeL##1{}\def\eqlabeL##1{}\def\reflabeL##1{}}
\def\writelabels{\def\wrlabeL##1{\leavevmode\vadjust{\rlap{\smash%
{\line{{\escapechar=` \hfill\rlap{\sevenrm\hskip.03in\string##1}}}}}}}%
\def\eqlabeL##1{{\escapechar-1\rlap{\sevenrm\hskip.05in\string##1}}}%
\def\reflabeL##1{\noexpand\llap{\noexpand\sevenrm\string\string\string##1}}}
\nolabels
\global\newcount\refno \global\refno=1
\newwrite\rfile
\def\defref{$^{{\hbox{\rfont [\the\refno]}}}$\nref}
\def\nref#1{\xdef#1{\the\refno}\writedef{#1\leftbracket#1}%
\ifnum\refno=1\immediate\openout\rfile=refs.tmp\fi
\global\advance\refno by1\chardef\wfile=\rfile\immediate
\write\rfile{\noexpand\item{#1\ }\reflabeL{#1\hskip.31in}\pctsign}\findarg}
\def\findarg#1#{\begingroup\obeylines\newlinechar=`\^^M\pass@rg}
{\obeylines\gdef\pass@rg#1{\writ@line\relax #1^^M\hbox{}^^M}%
\gdef\writ@line#1^^M{\expandafter\toks0\expandafter{\striprel@x #1}%
\edef\next{\the\toks0}\ifx\next\em@rk\let\next=\endgroup\else\ifx\next\empty%
\else\immediate\write\wfile{\the\toks0}\fi\let\next=\writ@line\fi\next\relax}}
\def\striprel@x#1{} \def\em@rk{\hbox{}} 
\def\lref{\begingroup\obeylines\lr@f}
\def\lr@f#1#2{\gdef#1{\defref#1{#2}}\endgroup\unskip}
\newwrite\lfile
{\escapechar-1\xdef\pctsign{\string\%}\xdef\leftbracket{\string\{}
\xdef\rightbracket{\string\}}}

\def\writestop{\def\writestoppt{\immediate\write\lfile{\string\p
ageno%
\the\pageno\string\startrefs\leftbracket\the\refno\rightbracket%
\string\def\string\secsym\leftbracket\secsym\rightbracket%
\string\secno\the\secno\string\meqno\the\meqno}\immediate\closeout\lfile}}
\def\writestoppt{}\def\writedef#1{}
\catcode`\@=12 

\helv

\centerline{\bold Lack of evidence for an odderon at small $t$}
\vskip 3mm
\centerline{A Donnachie}

\centerline{University of Manchester}
\vskip 3mm
\centerline{P V Landshoff}

\centerline{University of Cambridge}
\vskip 2mm

\helv

{\bold Abstract}

It is fundamental that the phase of an elastic scattering amplitude is
related to its energy variation. We repeat a previous fit to 
$pp$ and $p\bar p$ elastic scattering data from 
13 to 13000~GeV, taking better account of the very high accuracy of the 
13~TeV data. The conclusion remains that there is no evidence for the 
existence of an odderon in the small-$t$ data.

\vskip 10truemm

In an analysis of its highly-accurate 13~TeV elastic scattering data at small 
momentum transfer $t$, the TOTEM collaboration concluded\defref{\totemrho}{
TOTEM collaboration: G Antchev et al, Eur Phys J C79 (2019) 9, arXiv:1812.04732
}
that the ratio $\rho$ of the real to the imaginary part of the forward amplitude
was about 0.1. This is somewhat smaller than predictions from extrapolating from
its value at lower energies and the collaboration attributed this to the onset
of a significant contribution from odderon exchange. 

We disagreed with this analysis\defref\ad1
{A Donnachie and P V Landshoff, Phys Lett B 798 (2019) 135008, arXiv:1904.11218
}
and concluded that the value of $\rho$ at 13~TeV is rather close to 0.14, and 
that therefore there is no need for odderon exchange. Our main cricitism 
of the TOTEM analysis was that it ignored information linking the phase of i
the amplitude to its variation with energy, and that therefore it is not valid
to extract the value of $\rho$ from the 13~TeV data alone.

Our approach has in turn been criciticised by the collaboration\defref\newtotem
{
TOTEM collaboration: K Osterberg, https://arxiv.org/pdf/2202.03724.pdf
}
on the grounds that we did not take sufficient account of the very high 
accuracy of the 13~TeV data. This has led us to repeat our least-$\chi^2$ 
fit to all the $|t|<0.1$~GeV$^2$ $pp$ elastic scattering data from 13.76~GeV to
13~TeV, now including only the statistical errors in the 13~TeV data, where
previously we combined them in quadrature with the systematic errors. 

The effect on the fits to the lower-enery data is hardly noticeable, but
it gives a much more accurate fit to the 13~TeV data. The value for
$\rho$ is still close to 0.14.

For the 13~TeV data beyond the Coulomb peak, $|t|>0.02$~GeV$2$,
we obtain $\chi^2=0.75$ per data point. A similar calculation for
the 8~TeV data gives $\chi^2=0.06$ per data point. See figure 1.
We do not show the 7~TeV or 2.76~TeV data, for the reasons we explained 
before$^{[2]}$.

\topinsert{
\epsfxsize=0.5\hsize\epsfbox[65 65 400 295]{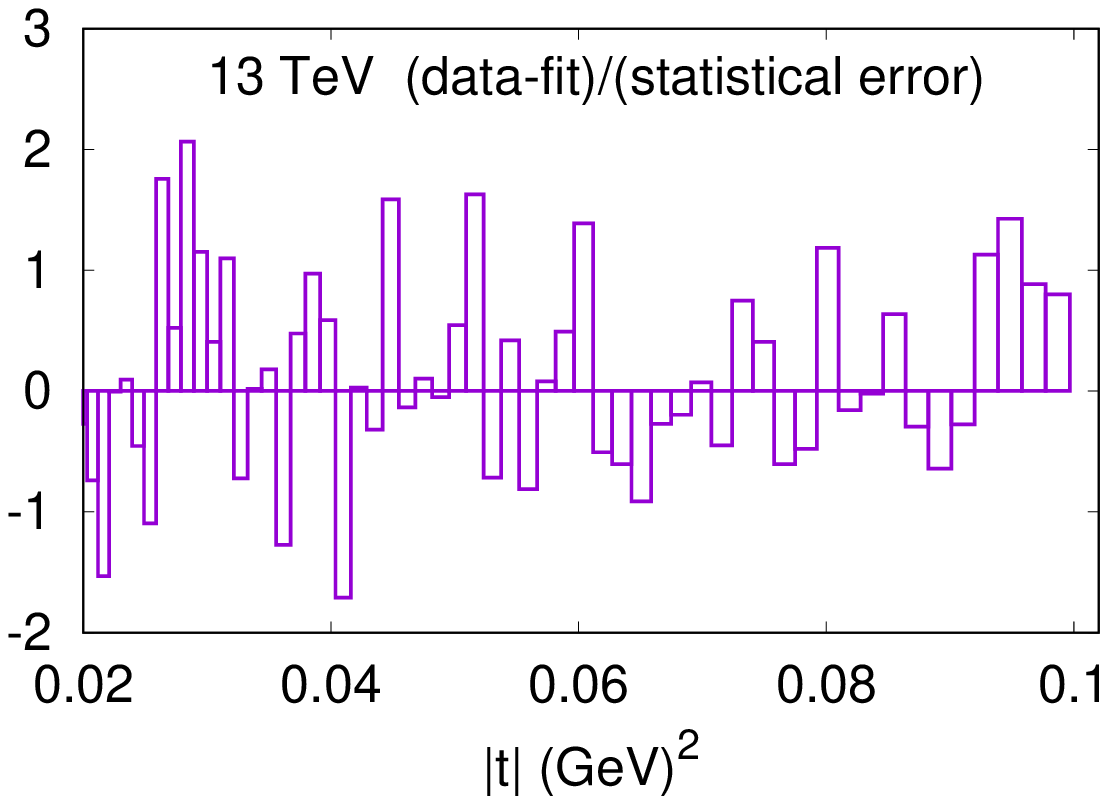}
\hfill
\epsfxsize=0.5\hsize\epsfbox[65 65 400 295]{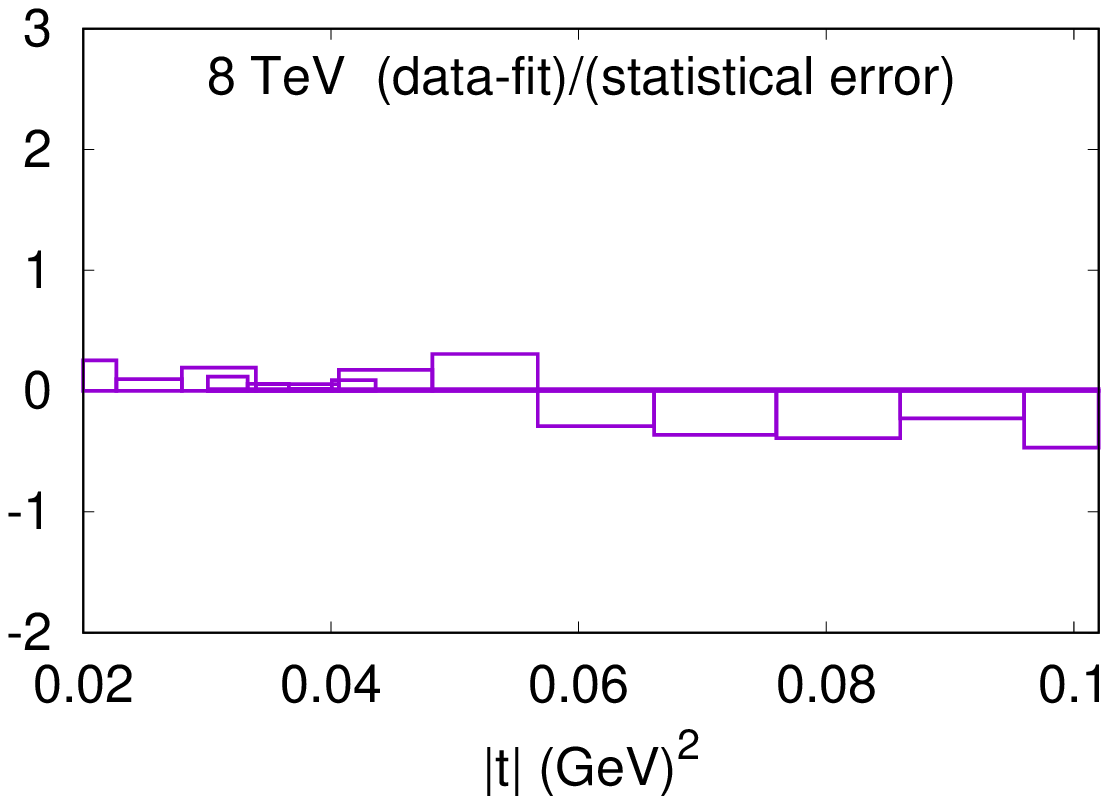}
\bigskip
\epsfxsize=0.5\hsize\epsfbox[60 60 320 290]{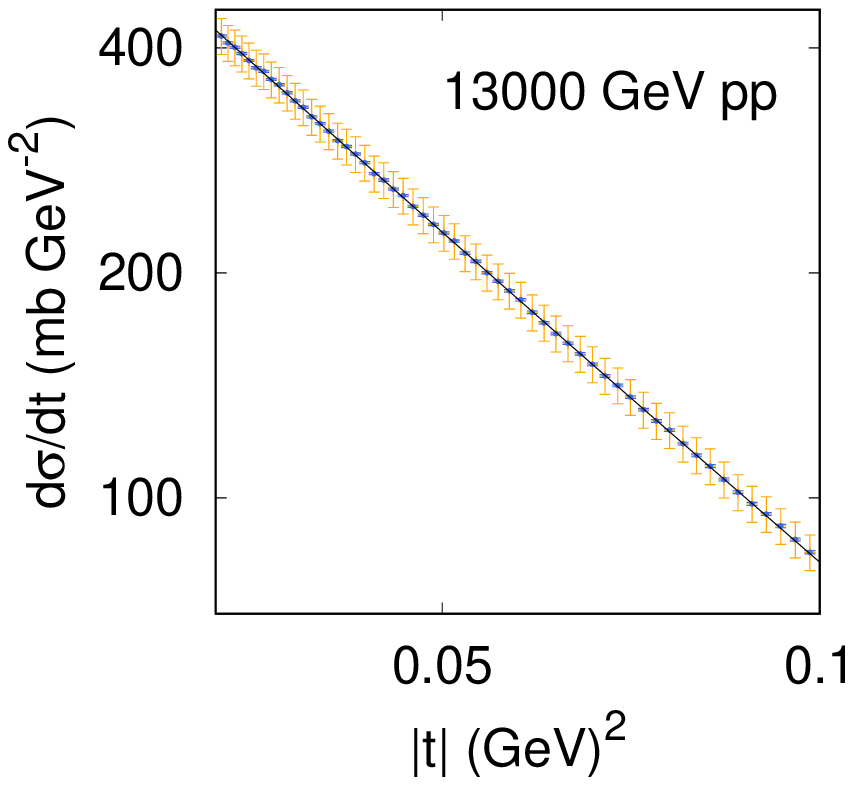}
\hfill
\epsfxsize=0.5\hsize\epsfbox[60 60 320 290]{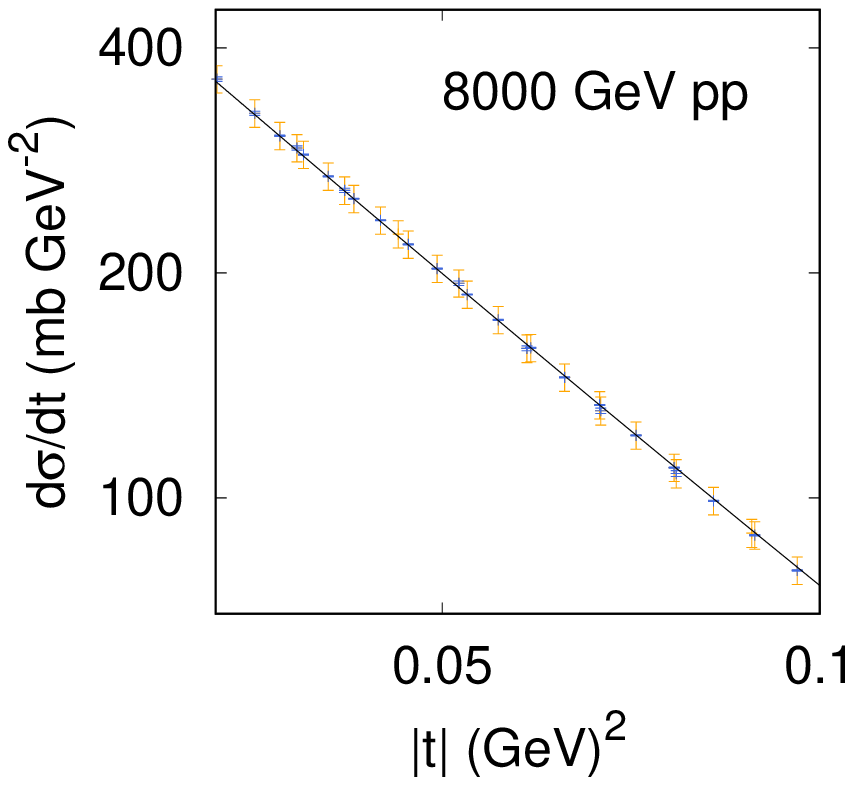}

Figure 1: fit to the 13 and 8~TeV data
\vskip 1 truecm
\epsfxsize=0.45\hsize\epsfbox[60 60 320 295]{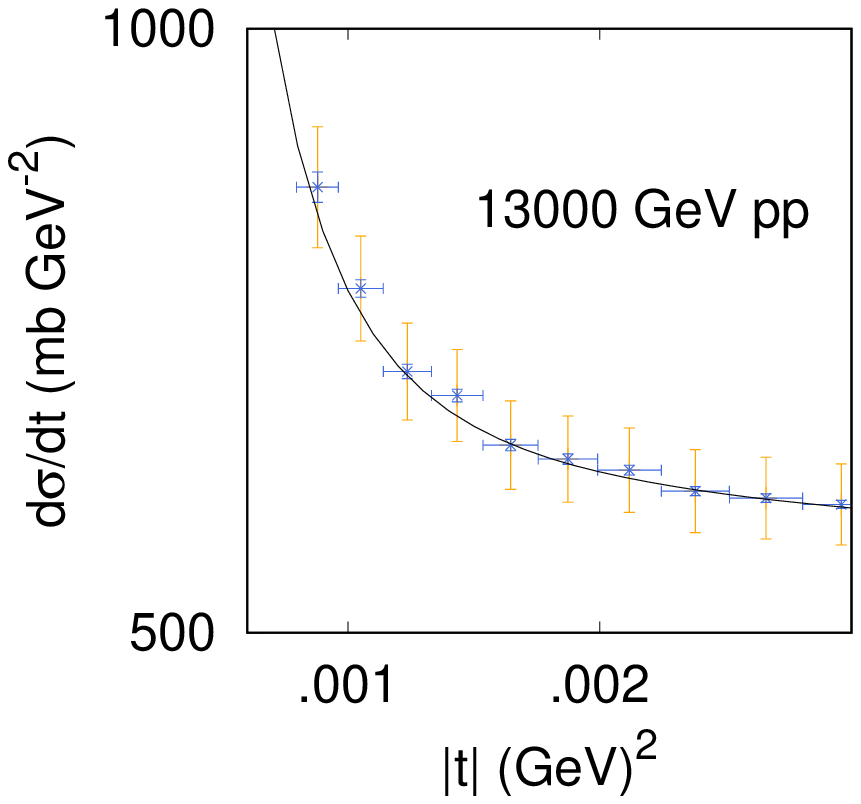}
\hfill
\epsfxsize=0.45\hsize\epsfbox[60 60 320 295]{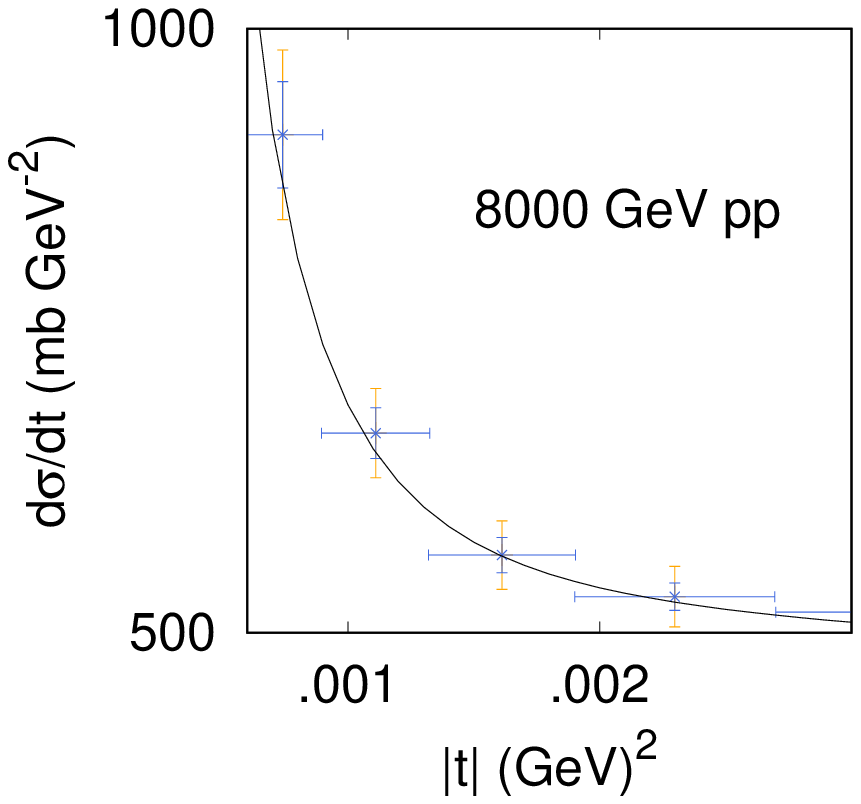}

Figure 2: fit to the 13 and 8~TeV data in the Coulomb peak
}
\endinsert

The fit is just as before$^{[2]}$, with the new parameter values
$$
\epsilon_{\P}=0.1083~~~  X_{\P}= 165.7~~~  X_+= 202.4~~~  X_-= 120.0~~~ 
\alpha'_{\P}=0.323\hbox{ GeV}^{-2}~~~ 
$$$$
A=0.594~~~  a_1=0.32\hbox{ GeV}^{-2}~~~ a_2= 8.192\hbox{ GeV}^{-2}
\eqno(1)
$$
with $C$ still fixed at 0.5.
Note that if we rounded the parameters to the accuracy we gave in our previous 
paper the values of $\chi^2$ quoted above would be very significantly worse.

\topinsert
{
\epsfxsize=0.45\hsize\epsfbox[60 60 320 295]{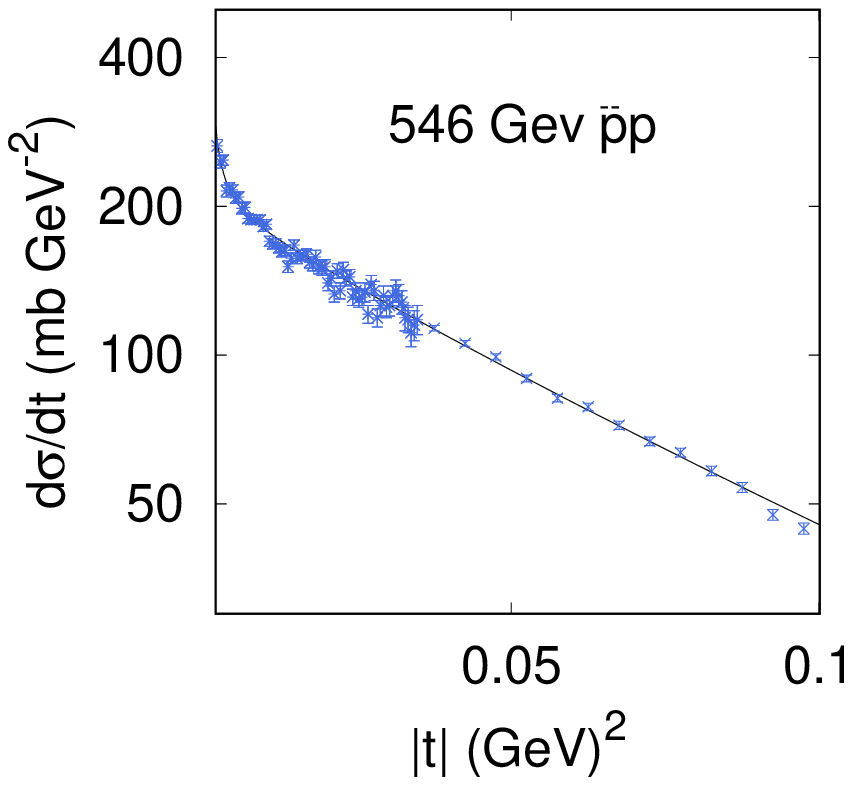}
\hfill
\epsfxsize=0.45\hsize\epsfbox[60 60 320 295]{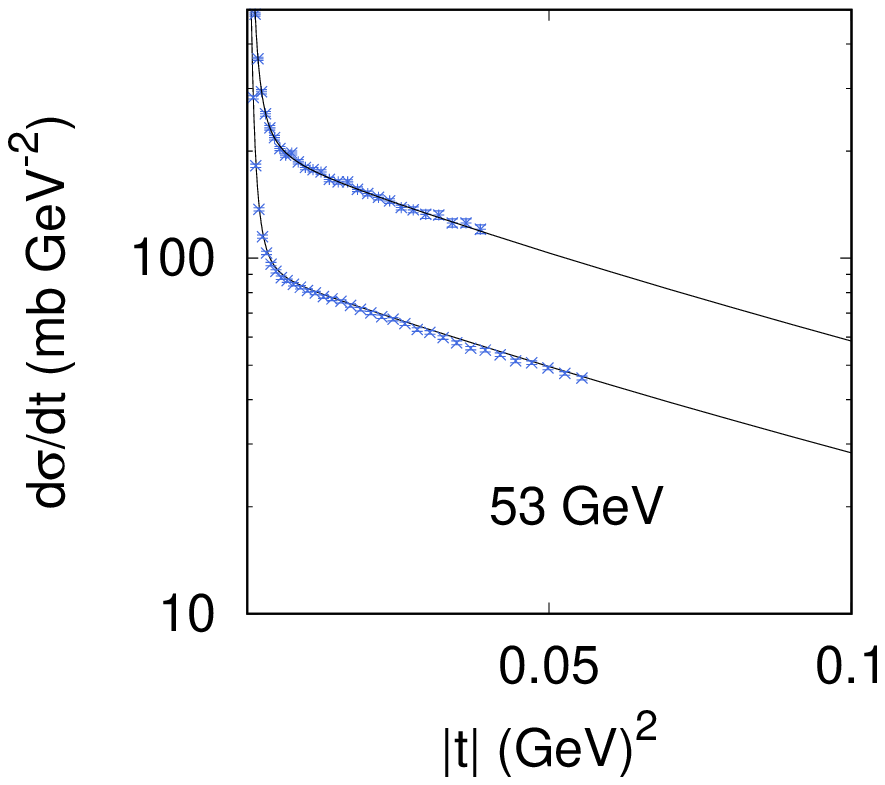}

Figure 3: Fits to lower-energy data. In the second plot the lower points are 
$pp$ elastic scattering, the upper points $p\bar p$ multiplied by 2.
\vskip 10truemm
\centerline{\epsfxsize=0.45\hsize\epsfbox[60 60 390 295]{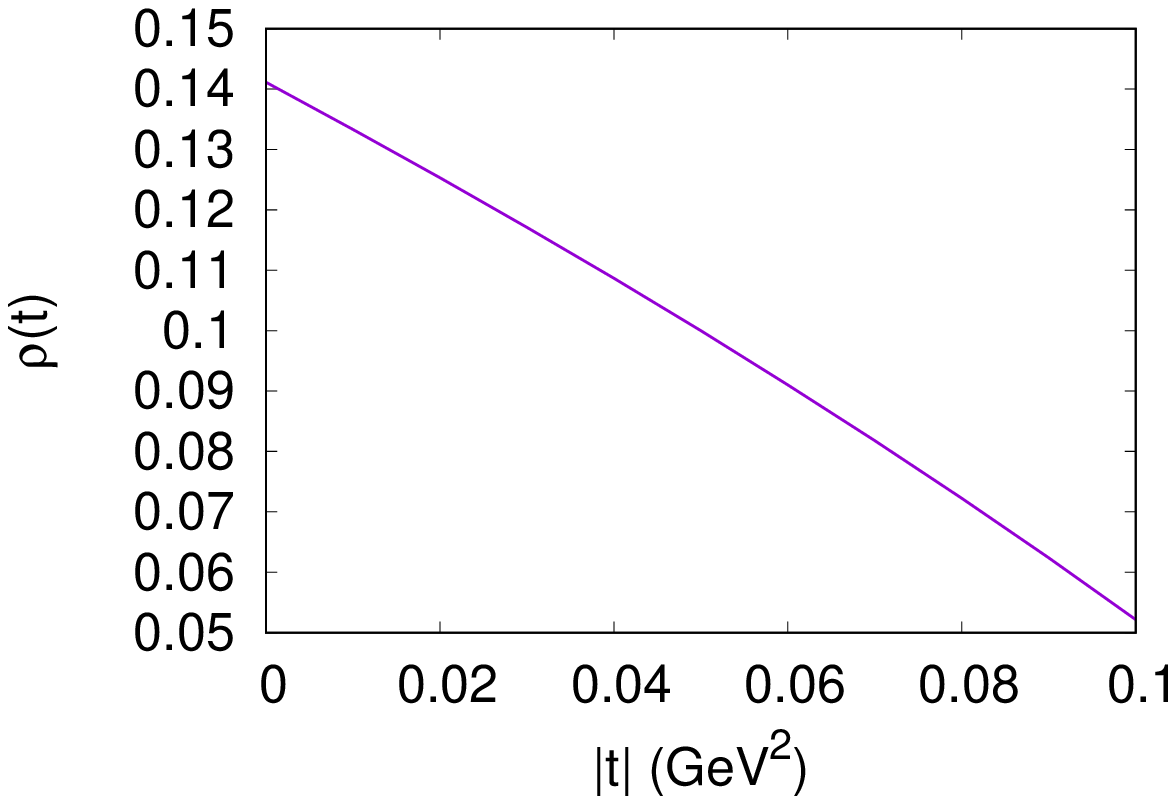}}
\bigskip
Figure 4: The calculated ratio $\rho(t)$ of the real to the imaginary 
part of the hadronic amplitude at 13~TeV.
}
\endinsert
Our fit used the data in the Coulomb peak and so it included in the $pp$
amplitude the term
$$
8\pi\alpha_{\hbox{\sevenrm EM}}F_1(t)^2/t
\eqno(2)
$$
where $F_1(t)$ is the Dirac form factor. Figure 2 shows that this
works quite well. It is the correct form to use at very small $t$, where
the difference from the more correct Rosenbluth form is negligible. 
We disagree with the claim by West and Yennie\defref\wy, 
{
G B West and D R Yennie, Physical Review 172 (1968) 1413
}
that including contributions from hadron exchange together with Coulomb exchange
introduces a phase factor, 
since if the photon is accompanied by any other exchange  
the result is not\defref\elop
{
R J Eden, P V Landshoff, D I Olive and J C Polkinghorne, {\oblique The
Analytic S-Matrix}, Cambridge University Press (1962) 
}
singular at $t=0$.

As we have said, the fits to the data at lower energies are almost as before.
Figure 3 shows two examples, at 546 and 53~TeV.

It is a matter of fundamental theory that if the variation with energy
of the amplitude varies with $t$, then so does its phase. Figure 4 shows that
the phase of the hadronic part of the amplitude at 13~TeV in the range
$0<|t|<0.1$ GeV$^2$ is nowhere near constant.

Of course a fit to the 13~TeV data alone would improve the already-good 
agreement between data and fit in the Coulomb region. With the simple
form (2) the value of $\rho$ reduces to a little over 0.12, while with the
West-Yennie modification to (2) it becomes less that 0.1, as TOTEM
found$^1$. However this destroys the agreement with the lower-energy
data and so is not a correct thing to do.

There is no need for an odderon at $t=0$. The data can be fitted perfectly
well without it.
\bigskip
\medskip\immediate\closeout\rfile\writestoppt
\baselineskip=0pt{{\bf References}}\h{\frenchspacing%
\parindent=20pt\escapechar=` \input refs.tmp\bigskip}\nonfrenchspacing
\bye